\documentclass[conference]{IEEEtran}  

\IEEEoverridecommandlockouts                              

\usepackage{graphicx} 
\usepackage{epsfig} 
\usepackage{amsmath} 
\usepackage{amssymb}  
\allowdisplaybreaks
\usepackage{multirow}
\usepackage[ruled, linesnumbered]{algorithm2e}

\let\labelindent\relax
\usepackage{enumitem}

\usepackage{float}
\usepackage{soul}
\usepackage{comment}
\usepackage[dvipsnames]{xcolor}
\usepackage{diagbox}

\usepackage{amsthm}

\theoremstyle{definition}

\theoremstyle{remark}

\usepackage{mathtools}
\newcommand{\ff}{\mathbf{f}}

\newcommand{\xx}{\mathbf{x}}
\newcommand{\uu}{\mathbf{u}}
\newcommand{\UU}{\mathbf{U}}

\newcommand{\bbe}{\mathbb{E}}

\newcommand{\bbp}{\mathbb{P}}
\newcommand{\bbr}{\mathbb{R}}

\begin{document}

\title{Eco-driving Accounting for Interactive \\ Cut-in Vehicles
}

\author{\IEEEauthorblockN{Chaozhe R. He}
\IEEEauthorblockA{\textit{Department of Mechanical and Aerospace Engineering} \\
\textit{University at Buffalo}\\
Buffalo, NY, U.S.A. \\
chaozheh@buffalo.edu}
\and
\IEEEauthorblockN{Nan Li}
\IEEEauthorblockA{\textit{Department of Aerospace Engineering} \\
\textit{Auburn University}\\
Auburn, AL, U.S.A. \\
nanli@auburn.edu}
}

\maketitle
\thispagestyle{plain}
\pagestyle{plain}

\begin{abstract}
Automated vehicles can gather information about surrounding traffic and plan safe and energy-efficient driving behavior, which is known as eco-driving. Conventional eco-driving designs only consider preceding vehicles in the same lane as the ego vehicle. In heavy traffic, however, vehicles in adjacent lanes may cut into the ego vehicle's lane, influencing the ego vehicle's eco-driving behavior and compromising the energy-saving performance. 
Therefore, in this paper, we propose an eco-driving design that accounts for neighbor vehicles that have cut-in intentions. Specifically, we integrate a leader-follower game to predict the interaction between the ego and the cut-in vehicles and a model-predictive controller for planning energy-efficient behavior for the automated ego vehicle. 
We show that the leader-follower game model can reasonably represent the interactive motion between the ego vehicle and the cut-in vehicle.
More importantly, we show that the proposed design can predict and react to neighbor vehicles' cut-in behaviors properly, leading to improved energy efficiency in cut-in scenarios compared to baseline designs that consider preceding vehicles only.
\end{abstract}


\section{Introduction}

Energy consumed by the transportation sector accounts for more than a quarter of the total energy consumed in the U.S. annually \cite{MonthlyEnergyReview2023Dec}, and improving energy efficiency carries great financial and societal benefits \cite{ardalan2020book}. 
Given the same vehicles, different driving profiles could result in great variations in energy consumption \cite{saltsman2014impacts}. 
Extensive research has been done on optimizing the control inputs (acceleration/deceleration, steering, etc) to achieve the best energy efficiency over given routes and/or driving scenarios, which is commonly referred to as eco-driving. 
Traditional eco-driving design uses geographic information to design optimal speed profiles, which can achieve more than 10\% reduction in energy consumption \cite{He2016Fuel, Sciarretta2015Review} in free-flow traffic. 

With the advances in automated vehicle technologies capable of more accurate perception of the surrounding environment, recently, researchers have been focusing on developing eco-driving controllers that consider the motions of surrounding vehicles in traffic, through reactive~\cite{chaozhe2020fuel}, predictive~\cite{borrelli2017ecoCACC}, and cooperative approaches \cite{2016platoon_review,shengbo2021distrEcoMPC,shengbo2018Urban, 2022CathyWu_EcoIntersection}. 
While many eco-driving designs bring significant energy benefits under various traffic scales, scenarios, and demand patterns, the designs and validations rarely consider lane change motions by the preceding vehicles. 
The eco-driving actions can lead to a large spacing between the ego vehicle and its preceding vehicles~\cite{Shen2023ReactiveVsPredictive}. In real-world traffic scenarios, this could lead to more frequent cut-in motions by vehicles from adjacent lanes, causing a big disturbance to the planned eco-driving behavior, and compromising the energy-efficiency performance~\cite{voronov2020cut}. 
The work of \cite{He2023EcoDrivingConsiderLaneChange} considered lane change motion in downstream traffic flow, but did not consider cut-in right in front of the ego.
The work of \cite{GU2022RLEcoDrivingMutliLane} used reinforcement learning to integrate longitudinal and lateral decision-making processes for eco-driving in mixed traffic scenarios. Despite performances achieved in long-duration simulations with various traffic conditions, it is not clear if the design handles cut-in vehicles efficiently. 
The work of \cite{LIANG2024Eco_driving_dynamic_preceding} considered cut-in~vehicles but the cut-in vehicles are not interactive. Given the interactive nature of cut-in motions, the cut-in vehicle's behavior may change in reaction to the ego's motion. 
Simply modeling the cut-in motion as non-reactive trajectories in an eco-driving design may not be effective when such interaction happens.

In this work, we propose an eco-driving controller that accounts for interactive cut-in vehicles. 
Specifically, the contributions are:
\begin{itemize}
    \item We use a game-theoretic approach to model the cut-in vehicle's behavior. The model generates different cut-in behaviors corresponding to different cut-in intentions of the vehicle while interacting with the traffic that it is cutting into. 
    \item We propose an eco-driving controller that considers interactive cut-in vehicles modeled using the game-theoretic approach. Based on estimations of the cut-in vehicle's intentions and corresponding predictions of its future motions, the controller plans for energy-efficient behavior for the ego vehicle. We show the benefits of considering cut-in vehicles in simulations where the cut-in vehicle is reacting to the ego's eco-driving behavior. 
\end{itemize}

\begin{figure}[t]
    \centering
    \includegraphics[width=0.47\textwidth]{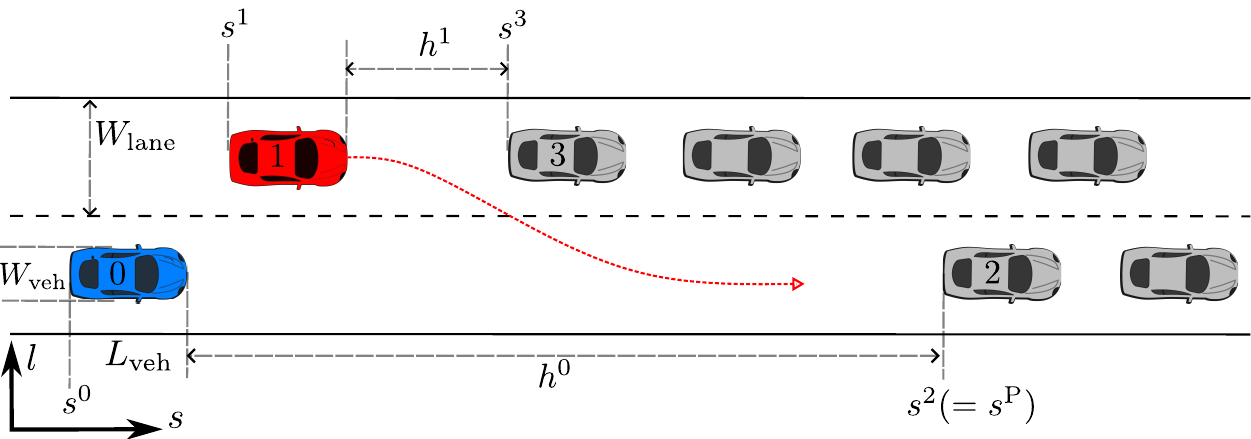}
    \caption{Top view of the scenario studied in this work. The automated ego vehicle (blue vehicle 0) is approaching slow traffic ahead (with grey vehicle 2 at the tail). In the lane on the left, there is also slow traffic building up (with grey vehicle 3 at the tail) and a target vehicle (red vehicle 1) is approaching this traffic and has the intention to cut into the ego vehicle's lane.}
    \label{fig:setup}
\vspace{-4mm}
\end{figure}

The remainder of this paper is organized as follows: Section~\ref{sec:modeling} introduces the problem setting and the dynamic models of vehicles. Section~\ref{sec:LFG} describes the leader-follower game-theoretic model used to model the interactive cut-in vehicle. Section~ \ref{sec:approach} presents the details of the proposed eco-driving controller that utilizes the game-theoretic model. Section~\ref{sec:results} evaluates the proposed design through simulation case studies. Section~\ref{sec:concl} concludes the paper and discusses future work.

\section{Problem Statement}\label{sec:modeling}

In this section, we first introduce the traffic scenario considered in this paper and then describe the models used to develop the eco-driving controller, including models representing the decision processes of all road vehicles and the ego vehicle's longitudinal dynamics that are suitable for an eco-driving controller design.

\subsection{Traffic Scenario of Interest}
The traffic scenario we consider is on a straight stretch of flat road with two lanes, illustrated in Fig.\,\ref{fig:setup}. The automated ego vehicle (blue vehicle 0) is approaching slow traffic ahead (with grey vehicle 2 at the tail). In the lane on the left, there is also slow traffic building up (with grey vehicle 3 at the tail) and a target vehicle (red vehicle 1) is approaching this traffic. 
Because the traffic in the left lane appears earlier in downstream and with the big gap between the ego and its preceding vehicles, the target vehicle may intend to cut into the ego vehicle's lane. 

The goal of this work is to design an eco-driving controller for the ego vehicle to approach the slow traffic ahead in an energy-efficient manner, while properly predicting and reacting to the cut-in vehicle's motion. 

\subsection{Traffic Dynamics}
We consider the following discrete-time equations of motion to describe vehicle kinematics during forward and lane change motions:
\begin{equation}\label{eqn:vehicle_kinematics}
    \begin{aligned}
        s(t+1) &= s(t) + v_{s}(t)\Delta t + \frac{1}{2} a_{s}(t) \Delta t^2, \\
        v_s(t+1) & = v_s(t) + a_{s}(t)\Delta t, \\
        l(t+1) &= l(t) + v_{l}(t)\Delta t,\\
    \end{aligned}
\end{equation}
where $s(t)$ and $l(t)$ denote the longitudinal and lateral positions of the vehicle at discrete time $t$; $v_{s}(t)$ and $v_{l}(t)$ denote the longitudinal and lateral speeds of the vehicle at\,$t$; $a_{s}(t)$ denotes the longitudinal acceleration of the vehicle at\,$t$; and $\Delta t$ is the sampling time interval. For vehicle $i$, we treat $\xx^{i} = [s^{i}, v^{i}_{s}, l^{i}]^{\top}$ as its state vector and $u^{i} = [a_{s}^{i}, v_{l}^{i}]^{\top}$ as its control input vector.

We focus on a scenario involving 4 vehicles, the ego vehicle~$0$, the potential cut-in vehicle $1$, and two other vehicles, $2$ and $3$, at the end of the slow traffic flow downstream; c.f., Fig.\,\ref{fig:setup}. For simplicity, we consider the interactive decision processes by the ego and potential cut-in vehicles, while assuming the other vehicles maintain their lanes and speeds, i.e., $u^{i} (t) = [0, 0]^{\top}$ for $i \geq 2$. 
Denote $\xx = [\xx^{0},\ldots, \xx^{3}]^{\top}$ and express the traffic dynamics in the following compact form:
\begin{equation}\label{eqn:compact_dynamics}
    \begin{aligned}
            \xx(t+1) &= \ff(\xx(t), u^{0}(t), u^{1}(t)) = \begin{bmatrix}
            f(\xx^{0}(t), u^{0}(t)) \\
            f(\xx^{1}(t), u^{1}(t)) \\
            f(\xx^{i}(t), u^{i}(t))
        \end{bmatrix},
    \end{aligned}
\end{equation}
where $f$ is given by the equations in \eqref{eqn:vehicle_kinematics}.



\subsection{Longitudinal Vehicle Dynamics}

\begin{figure}
    \centering
    \includegraphics[width=0.48\textwidth]{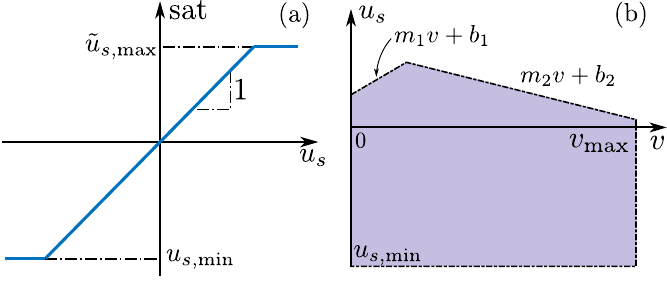}
    \caption{Nonlinear functions in the vehicle dynamics. (a) Saturation function\,\eqref{eqn:sat_def}. (b) Acceleration limits\,\eqref{eqn:umax_def}.}
    \label{fig:sat_fun}
    \vspace{-4mm}
\end{figure}

In this work, we incorporate a higher-fidelity longitudinal vehicle dynamics model and a powertrain model for the ego vehicle. They are useful for incorporating practical powertrain limits and evaluating energy consumption, which are important for an eco-driving design. Assuming the vehicle is driving on a flat road, the longitudinal acceleration is given by the following equation~\cite{orosz2020survey}:
\begin{equation}\label{eqn:veh_dyn_original}
a_{s}(t) = -\frac{1}{m_{\mathrm{eff}}}\Big(m{\rm g}\zeta + kv_{s}(t)^2\Big) + \frac{T_{\mathrm{w}}(t)}{m_{\mathrm{eff}}R},
\end{equation}
where the effective mass ${m_{\mathrm{eff}} = m + I/R^2}$ incorporates the vehicle mass $m$, mass moment of inertia $I$, and the radius $R$ of the wheels; $\text{g}$ denotes the gravitational constant, $\zeta$~denotes the rolling resistance coefficient, and $k$ denotes the air resistance coefficient. The acceleration is determined by the torque $T_{\mathrm{w}}$ on the wheels delivered by a powertrain equipped with engine/electric motors and brakes. We model the powertrain as one that takes a control command $u_{s}$ in the scale of acceleration and applies proper scaling to deliver the corresponding torque $T_{\rm w}$ but with delay and speed-dependent saturation, i.e.,
\begin{equation}\label{eqn:powertrain_model}
    T_{\rm w}(t) = m_{\mathrm{eff}}R \, \mathrm{sat}\big(u_{s}(t-\iota), v_{s}(t)\big).
\end{equation}
The speed-dependent saturation $\mathrm{sat}(\cdot, v)$ arises from engine/motor power and torque limits and braking capability. It is modeled as
\begin{subequations}\label{eqn:sat_0}
\begin{align}
\mathrm{sat}(u_{s}, v_{s}) &= \min\left\{\max\{u_{s},u_{s,\min}\},\tilde{u}_{s, \max}(v_{s})\right\}, \label{eqn:sat_def} \\
\tilde{u}_{s, \max}(v_{s}) &= \min\left\{u_{s, \max}, m_1 v_{s} + b_1, m_2 v_{s} + b_2\right\}, \label{eqn:umax_def}
\end{align}
\end{subequations}
as illustrated in Fig.\,\ref{fig:sat_fun}(a) and (b). In \eqref{eqn:sat_0}, $u_{s, \min}$ represents the minimum acceleration\,(i.e., maximum deceleration) due to the braking capability, and $u_{s, \max}$, $m_1$, $m_2$, $b_1$, $b_2$ are parameters determined by engine/motor power and torque limits. In our torque model \eqref{eqn:powertrain_model}--\eqref{eqn:sat_0}, the delay $\iota$ represents the time gap between a command being sent to the powertrain and being executed, while the saturation dependence on speed is modeled as the final lumped effect and hence not subject to the delay~$\iota$.

Combining \eqref{eqn:veh_dyn_original} and \eqref{eqn:powertrain_model}, we obtain
\begin{equation}\label{eqn:simple_nonlin_dyn}
a_{s}(t) = -\varrho\big(v_{s}(t)\big) + \mathrm{sat}\big(u_{s}(t-\iota), v_{s}(t)\big),
\end{equation}
where
\begin{equation}\label{eqn:f_def}
\varrho(v_{s}) = \frac{1}{m_{\mathrm{eff}}}\left(m{\rm g}\zeta + kv_{s}^2\right).
\end{equation}
Thus, for the vehicle to apply a desired acceleration $a_{\mathrm{d}}$, one needs to send a control command $u_{s}$ that compensates the resistance force, i.e., 
\begin{equation}\label{eqn:ucomp_ftil}
u_{s}(t) = \varrho\big(v_{s}(t)\big) + a_{\mathrm{d}}(t).
\end{equation}
However, the compensation term ${\varrho}\big(v_{s}(t)\big)$ will also be delayed by $\iota$ and thus cannot provide perfect compensation. This fact is accounted for in our simulations.

\subsection{Cut-in Vehicle Decision Model}
We model the cut-in vehicle as an interactive agent that makes decisions accounting for the surrounding vehicles' reactions (including the ego vehicle). We assume that it takes high-level actions from a finite set to maximize a cumulative reward over a horizon based on its prediction of traffic dynamics.

\subsubsection{Action space}
We assume that vehicles take high-level actions from the following set $A$: 
\begin{itemize}
    \item \textit{Maintain}: maintain current speed and lateral position.
    \item \textit{Mildly accelerate}: maintain the current lateral position and accelerate at $\Delta a_{\rm mild}$ while staying below velocity upper limit $v_{\max}$.
    \item \textit{Mildly decelerate:} maintain the current lateral position and decelerate at $-\Delta a_{\rm mild}$ while staying above velocity lower limit $v_{\min}$.
    \item \textit{Hard accelerate}: maintain the current lateral position and accelerate at $\Delta a_{\rm hard}$, with $\Delta a_{\rm hard} > \Delta a_{\rm mild}$, while staying below velocity upper limit $v_{\max}$.
    \item \textit{Hard decelerate}: maintain the current lateral position and decelerate at $-\Delta a_{\rm hard}$ while staying above velocity lower limit $v_{\min}$.
    \item \textit{Steer to left}: move towards left with lateral velocity of $v_{l}=\frac{W_{\rm lane}}{2}$, where $W_{\rm lane}$ is the lane width.
    \item \textit{Steer to right}: move towards right with lateral velocity of $v_{l}=-\frac{W_{\rm lane}}{2}$.
\end{itemize}
Based on the above actions, a continuous lane change/cut-in takes $2$ seconds to complete.

\subsubsection{Reward}

The reward function is given as
\begin{equation}\label{eqn:game_running_reward}
    R(\xx, u^{\rm self}, u^{\rm other}) = \boldsymbol{\omega}^{\top}\mathbf{r},
\end{equation}
where $\mathbf{r} = [r_{1},\ldots, r_{6}]^{\top}$ contains reward terms and $\boldsymbol{\omega} \in \bbr_{+}^{6}$ is the vector of weights; $u^{\rm self}$ is the action taken by the vehicle itself (e.g., the ego vehicle), while $u^{\rm other}$ is the action taken by the vehicle that it is interacting with (e.g., the cut-in vehicle). The reward terms are defined as follows:
\begin{itemize}
    \item $r_{\rm 1} \in \{-1, 0\}$ is an indicator for vehicle collisions. Each vehicle is represented by a rectangle bounding box. If the ego vehicle's bounding box, defined by the length and width of the vehicle, $L_{\rm veh} \times W_{\rm veh}$, overlaps with that of any other vehicle in the traffic, then $r_{1} = -1$, and $r_{1} = 0$ otherwise. The weight $\omega_{1}$ for $r_{1}$ is chosen to be large enough to prioritize collision avoidance.
    \item $r_{\rm 2} \in \{-1, 0\}$ indicates if the vehicle is getting too close to its preceding vehicles
    \begin{equation}
        r_{\rm 2} = 
        \begin{cases}
            -1 & \text{if}\,\, h^{\rm self} < v^{\rm self}_s\, \tau_{\rm desired},\\
            0 & \text{otherwise},
        \end{cases}
    \end{equation}
    where $h^{\rm self} = s^{\rm P} - s^{\rm self} - L_{\rm veh}$ is the distance headway to the preceding vehicle, $v^{\rm self}_s$ is the longitudinal speed, and $\tau_{\rm desired}$ is the desired time headway. 
    \item $r_{\rm 3} = s$ defines distance liveness, which gives the motivation for performing a lane change/cut-in: the vehicle chooses to change lanes if it leads to traveling a longer distance over a certain time window. 
    \item $r_{\rm 4} = \frac{v^{\rm self}_s - v_{\max}}{v_{\max}}$ defines speed liveness,  which gives the incentive to reach the maximum speed if possible.
    \item $r_{\rm 5} = -\, |l^{\rm self} - l_{\rm target}|$ defines lateral liveness, which gives the incentive to complete a lane change.
    \item $r_{\rm 6}$ penalizes control effort and is equal to the negative of the norm of $u^{\rm self}$.
\end{itemize}

During a lane change, the preceding vehicle changes. The preceding vehicle is determined by the following rule:
\begin{equation}\label{eqn:determine_preceding_vehicle}
\begin{aligned}
    {{\rm P}} \in \underset{i \in \mathcal{I}}{\arg\min} \quad & s^{i} - s^{\rm self} - L_{\rm veh}, \\
     \rm {s.t.} \quad & s^{i} - s^{\rm self} - L_{\rm veh} \geq 0, \\
                      &  |l^{i} - l^{\rm self}| \leq W_{\rm veh}, \\    
\end{aligned}
\end{equation}
where the index $i$ iterates through $\mathcal{I}$ which is the set of all surrounding vehicles whose longitudinal and lateral positions are $s^{i}$ and $l^{i}$, respectively.


We assume that the cut-in vehicle plans its motion using a receding-horizon optimization approach: At each time $t$, the vehicle calculates an optimal control sequence $\mathbf{u}^{\ast}(t) = \{u^{\ast}(t), u^{\ast}(t+1),\ldots, u^{\ast}(t + N - 1)\}$ that maximizes the cumulative reward over the planning horizon, i.e.,
\begin{equation}\label{eqn:cum_reward}
    \begin{aligned}
    &\uu^{{\rm self}, \ast}(t) \in \arg\max \\
    &\sum_{k = 0} ^{N-1} \lambda^{k} R(\xx(k+1|t), u^{\rm self}(k|t), u^{\rm other}(k|t)),
    \end{aligned}
\end{equation}
where $\lambda \in (0,1]$ is a discount factor. Note that such a decision process by the cut-in vehicle requires prediction of the interacting vehicle's (i.e., the ego vehicle's) actions, $\uu^{\rm other}(t)$. We use a game-theoretic approach for this prediction, introduced in Section\,\ref{sec:LFG}.  

\section{Leader-Follower Game-Theoretic Model for Interactive Cut-in Vehicles}\label{sec:LFG}

During a cut-in process, the vehicle performing the lane change, referred to as the cut-in vehicle, interacts with the vehicle driving in the target lane, which may yield to the cut-in vehicle or proceed without regard to the cut-in vehicle's intention. In this work, we consider a leader-follower game-theoretic model to represent the drivers' interaction intentions and the resulting vehicle behaviors. In this model, a driver/vehicle can take a leader or a follower role which will lead to different decision strategies: a follower considers all possible behaviors by the leader and makes the best decision against worst-case outcomes, while a leader makes the best decision assuming the other using the follower's decision strategy. This leader-follower game-theoretic model has shown promise in modeling various vehicle interaction scenarios, including intersections \cite{Li2020GameOriginal}, highway forced merge \cite{Liu2023}, and highway overtaking \cite{JI2023GameOvertake}.

Consider a pair of self and other vehicles to be a pair of leader and follower, and rewrite the cumulative reward in \eqref{eqn:cum_reward}~as
\begin{equation}
\begin{aligned}
\bar{R}_{\sigma}&\left(\xx(t), \uu_{\rm l}(t), \uu_{\rm f}(t)\right) = \\
&\sum_{k = 0}^{N-1} \lambda^{k}R_{\sigma}\left(\xx(k + 1|t), u_{\rm l}(k|t), u_{\rm f}(k|t)\right),
\end{aligned}
\end{equation}
where $\uu_{\rm l}(t) = \{u_{\rm l}(k|t)\}_{k = 0}^{N-1} \in \UU_{\rm l} = A^N$ and $\uu_{\rm f}(t) = \{u_{\rm f}(k|t)\}_{k = 0}^{N-1} \in \UU_{\rm f} = A^N$ denote the action sequences of the leader and the follower, the subscript $\sigma \in \{ \rm leader (l), follower (f)\}$ represents the role in the game, and ${R}_{\sigma}(\cdot, \cdot, \cdot)$ is the single-step reward defined in \eqref{eqn:game_running_reward}, with subscript $\sigma$ implying that the reward is calculated for vehicle of different roles.

The leader and the follower both attempt to maximize their cumulative rewards but follow different strategies: The\,follower maximizes the worst-case reward due to uncertain leader's actions, i.e., it takes the following ``max-min'' strategy:
\begin{subequations}\label{eqn:follower_decision_strategy}
\begin{equation}
\uu^{\ast}_{\rm f}(t) \in \underset{\uu_{\rm f} \in \UU_{\rm f}}{\arg\max}\, Q_{\rm f}(\xx(t), \uu_{\rm f}),
\end{equation}
where
\begin{equation}
Q_{\rm f}(\xx(t), \uu_{\rm f}) = \min_{\uu_{\rm l} \in \UU_{\rm l}} \bar{R}_{\rm f}(\xx(t), \uu_{\rm l}, \uu_{\rm f}).
\end{equation}
\end{subequations}
This strategy represents a ``cautious'' or ``conservative'' driving strategy, or an intention to yield \cite{Liu2023}. In contrast, the leader assumes that the other vehicle is a follower and hence uses the above ``max-min'' strategy. Therefore, the leader can predict the follower's actions and takes its own actions according to:
\begin{subequations}\label{eqn:leader_decision_strategy}
\begin{equation}
\uu^{\ast}_{\rm l}(t) \in \underset{\uu_{\rm l} \in \UU_{\rm l}}{\arg\max}\, Q_{\rm l}(\xx(t), \uu_{\rm l}),
\end{equation}
where
\begin{align}
& Q_{\rm l}(\xx(t), \uu_{\rm l}) = \min_{\uu_{\rm f} \in \UU_{\rm f}^{\ast}(\xx(t))} \bar{R}_{\rm l}(\xx(t), \uu_{\rm l}, \uu_{\rm f}), \\
& \UU_{\rm f}^{\ast}(\xx(t)) = \{ \uu_{\rm f} \in \UU_{\rm f}: Q_{\rm f}(\xx(t), \uu_{\rm f}) \geq Q_{\rm f}(\xx(t), \uu_{\rm f}'), \nonumber \\
&\quad\quad\quad\quad\quad\quad\quad\quad\quad\quad\quad\quad\quad\quad \forall \uu_{\rm f}' \in \UU_{\rm f}\}.
\end{align}
\end{subequations}
This strategy represents a driver/vehicle that assumes the other vehicle will yield and hence decides to proceed more aggressively \cite{Liu2023}.

The leader-follower game-theoretic model is suitable for modeling a cut-in vehicle's interactive behavior for two reasons:
First, the asymmetry between the leader's and the follower's decision processes \eqref{eqn:follower_decision_strategy} and \eqref{eqn:leader_decision_strategy} can be used to model different cut-in behaviors -- for example, a cut-in in front of the ego vehicle versus behind the ego vehicle, or directly cut-in versus cut-in after speed up. Second, the optimal decisions depend on the traffic state -- the optimal action sequences change as the ego vehicle's behavior changes.


\section{Eco-Driving Controller that Accounts for Cut-in Vehicles}\label{sec:approach}

In this section, we describe the eco-driving controller that accounts for cut-in vehicles. 

\subsection{MPC-based Eco-driving Controller}

We assume that the ego vehicle does not change lanes and maintains its lateral position $l^{0}(t)$ constant. Denote the ego vehicle's longitudinal state at time $t$ as $x^{0}(t) = [s^{0}(t), v_{s}^{0}(t)]^{\top}$ and its preceding vehicle's longitudinal state at time $t$ as $x^{\rm P}(t) = [s^{\rm P}(t), v_{s}^{\rm P}(t)]^{\top}$. The eco-driving controller determines control input based on the following optimization problem:
\begin{equation}\label{eqn:mpc_continuous}
\small
\begin{aligned}
\min \, & \underset{\bbp\left(\hat{x}^{\rm P}(\cdot|t)\right)}{\bbe}\left\{\int_{0}^{T} \ell \left(x^{0}\big(\tilde{t}|t\big), \hat{x}^{\rm P}\big(\tilde{t}|t\big), a_{s}^{0}\big(\tilde{t}|t\big) \right) \mathrm{d}\tilde{t}\,\right\},
\\
\mathrm{s.t.}\, &\, G_{\mathrm{dynamics}}\Big( x^{0}\big(\tilde{t}|t\big), a_{s}^{0}\big(\tilde{t}|t\big) \Big) = 0,
\\
&\, G_{\mathrm{saturation}}\Big( x^{0}\big(\tilde{t}|t\big), a_{s}^{0}\big(\tilde{t}|t\big) \Big) \leq 0, 
\\
&\, \bbp\left\{G_{\mathrm{safety}}\Big( x^{0}\big(\tilde{t}|t\big), \hat{x}^{\rm P}\big(\tilde{t}|t\big) \Big) \leq 0, \forall \tilde{t}\in (0, T] \right\} \geq 1 - \eta, 
\end{aligned}
\end{equation}
where $x^{0}(t|t) = x^{0}(t)$ is the ego vehicle's current state, and $a^{0}_{s}\big(\tilde{t}|t\big) = a_{\text{d}}^{0}\big(t+\tilde{t}-\iota\big)$ for $\tilde{t}\in [0, \iota)$ due to the powertrain delay. The eco-driving controller aims to minimize the cumulative cost over the time horizon $[0, T]$ subject to vehicle dynamics constraints $G_{\mathrm{dynamics}}$, powertrain saturation constraints $G_{\mathrm{saturation}}$, and certain safety constraints $G_{\mathrm{safety}}$ based on predictions of the ego vehicle's state $x^{0}\big(\tilde{t}|t\big)$ and the preceding vehicle's state $\hat{x}^{\rm P}\big(\tilde{t}|t\big)$. In particular, the cost function $\ell$ is designed for the vehicle to track a desired speed-dependent car-following distance, given in~\eqref{eqn:constant_time_headway}, while penalizing energy consumption due to control effort:
\begin{equation}\label{eqn:constant_time_headway}
H(v_{s}^{0}) = d + \tau v_{s}^{0},
\end{equation}
where $\tau$ represents a desired constant time headway. We consider minimizing the expected value of the cumulative cost and enforcing the safety constraints up to a prescribed probability level $1 - \eta$. This is because due to the cut-in vehicle, the preceding vehicle may change, according to \eqref{eqn:determine_preceding_vehicle}, and hence the preceding vehicle's state $\hat{x}^{\rm P}\big(\tilde{t}|t\big)$ over the horizon is uncertain. Therefore, we consider stochastic predictions, i.e., $\hat{x}^{\rm P}\big(\tilde{t}|t\big)$ is random and follows a certain distribution. We will elaborate on the stochastic prediction of $\hat{x}^{\rm P}\big(\tilde{t}|t\big)$ later in this~section.

We convert and solve the optimization problem \eqref{eqn:mpc_continuous} in discrete time. The vehicle dynamics constraints $G_{\mathrm{dynamics}}$ in discrete time are given by \eqref{eqn:vehicle_kinematics}, the powertrain saturation constraints $G_{\mathrm{saturation}}$ are defined according to \eqref{eqn:veh_dyn_original}--\eqref{eqn:sat_0}, plus a saturation on longitudinal speed that prevents the vehicle from speeding:
\begin{equation}\label{eqn:saturation_v}
    0 \leq v_{s}^{0}(k|t) \leq v_{\max},
\end{equation}
for $k = 1,\ldots, N$.

We define the safety constraints $G_{\mathrm{safety}}$ to enforce the predicted car-following distance,
\begin{equation}\label{eqn:headway_prediction}
\hat{h}^{0}(k|t) = \hat{s}^{\rm P}(k|t) - s^{0}(k|t) - L_{\rm veh},
\end{equation}
to be greater than a minimum distance,
\begin{equation}\label{eqn:minimal_distance}
H_{\min}(v_{s}^{0}) = d_{\min} + \tau_{\min} v_{s}^{0},
\end{equation}
to guarantee collision avoidance. Specifically, to compensate for prediction inaccuracy, we impose an additional safety margin $d_{\mathrm{margin}}(k)$ at each time $k$ over the prediction horizon. This leads to the following safety constraints:
\begin{equation}\label{eqn:safety_ineq}
\hat{h}^{0}(k|t) - H_{\min}\big(v_{s}^{0}(k|t)\big) - d_{\mathrm{margin}}(k) \geq 0,
\end{equation}
for $k = 1,\ldots, N$. The safety margin $d_{\mathrm{margin}}(k)$ is elaborated in \cite{Shen2023ReactiveVsPredictive} and for a given confidence level its value depends on~$k$ and is independent of the state $x^{0}(k|t)$.

We now explain how the preceding vehicle's state over the horizon, $\hat{x}^{\rm P}(k|t)$, relates to the cut-in vehicle's leader/follower role in its interaction with the ego vehicle. Recall that according to our leader-follower game-theoretic model, for a certain role of the cut-in vehicle, either leader or follower, we can predict its actions deterministically using \eqref{eqn:follower_decision_strategy} or \eqref{eqn:leader_decision_strategy}. If it is predicted that this vehicle will cut in front of the ego vehicle, this cut-in vehicle will become the preceding vehicle according to \eqref{eqn:determine_preceding_vehicle}. Correspondingly, $\hat{x}^{\rm P}(k|t)$ will be determined by the state of the cut-in vehicle. That is, if the role of the cut-in vehicle is given, we can predict $\hat{x}^{\rm P}(k|t)$ deterministically (the procedure of which will be elaborated in Section~\ref{sec:prediction_accouts_for_cut_in}). However, in a real-world cut-in scenario, the role of the cut-in vehicle is typically not known a priori. Instead, we assume we know a probability distribution of its role: $\sigma \in \{ \rm l, f\}$, $\sigma \sim \bbp(\sigma)$. Then, the distribution of the preceding vehicle's state $\hat{x}^{\rm P}(k|t)$ is determined by the probability distribution of the leader/follower role, $\sigma \sim \bbp(\sigma)$. We will describe how to estimate the distribution $\sigma \sim \bbp(\sigma)$ using online data in~Section~\ref{sec:LFG_role_estimation}.

In summary, the eco-driving controller solves the following discrete-time optimization problem to determine control: 
\begin{align*}
\min_{\substack{a_{s}^{0}(0|t), \\ \ldots \\ a_{s}^{0}(N-1+q|t)}} \!\!\!& \underset{\bbp\left(\hat{s}^{\rm P}(\cdot|t)\right)}{\bbe}\Bigg\{q_{\mathrm{g}} \sum_{k=1}^{N} \left(\hat{h}^{0}(k|t) - H\big(v_{s}^{0}(k|t)\big)\right)^2 
\\[-10pt] 
&\hspace{40pt} + q_{\mathrm{a}} \sum_{k=0}^{N-1} \left(a_{s}^{0}(k|t)\right)^2 \Bigg\}
\\
\mathrm{s.t.} \quad & s^{0}(k+1|t) = s^{0}(k|t) \!+\! v^{0}_{s}(k|t)\Delta t  \!+\! \frac{1}{2} a^{0}_{s}(k|t)\Delta t^2,
\\
& v^{0}_{s}(k+1|t) = v^{0}_{s}(k|t) \!+\! a^{0}_{s}(k|t)\Delta t,
\\
& \hspace{110pt} \forall\, k = 0, \ldots, N-1, 
\\
& 0\leq v^{0}_{s}(k|t) \leq v_{\max}\,, \quad \forall \, k = 1, \ldots, N,  
\\
& u_{\min}\leq a_{s}^{0}(k + q|t),
\\
& a_{s}^{0}(k+q|t) \leq m_1 v_{s}^{0}(k+q|t) + b_1,
\\
& a_{s}^{0}(k+q|t) \leq m_2 v_{s}^{0}(k+q|t) + b_2, \\
& \hspace{120pt} \forall\, k=0,\ldots, N-1, 
\\
& \hat{h}^{0}(k|t) = \hat{s}^{\rm P}(k|t) - s^{0}(k|t) - L_{\rm veh},
\\
& \bbp\bigg\{\hat{h}^{0}(k|t) - H_{\min}\big(v_{s}^{0}(k|t)\big) - d_{\mathrm{margin}}(k) \geq 0, 
\\
& \hspace{70pt}\forall\, k = 1, \ldots, N \bigg\} \geq 1 -\eta,
\\
&\hat{s}^{\rm P}(k|t) = \hat{s}_{\sigma}^{\rm P}(k|t), \quad \sigma \sim \bbp(\sigma), 
\\
&s^{0}(0|t) = s^{0}(t), \,\, v_{s}^{0}(0|t) = v_{s}^{0}(t),
\\
& a_{s}^{0}(k|t) = a_{s}^{0}(t + k - q), \quad q = \frac{\iota}{\Delta t},\\
&  \hspace{120pt} \forall \, k = 0, \ldots, q-1. \stepcounter{equation}\tag{\theequation}\label{eqn:mpc_discrete}
\end{align*}
At time $t$, eco-driving controller solves \eqref{eqn:mpc_discrete} for sequence $\{a_{s}^{0}(0|t),\ldots, a_{s}^{0}(N-1+q|t)\}$, and apply $a_{\rm d}^{0}(t) = a_{s}^{0}(q|t)$ to \eqref{eqn:simple_nonlin_dyn} using \eqref{eqn:ucomp_ftil} to compensate powertrain delay $\iota$. 


\subsection{Estimation of Cut-in Vehicle's Intention}\label{sec:LFG_role_estimation}
Here we describe the method for estimating the cut-in vehicle's intended role. From the perspective of the ego vehicle, the traffic dynamics \eqref{eqn:compact_dynamics} evolve with the cut-in vehicle~$1$ taking actions according to either \eqref{eqn:follower_decision_strategy} or \eqref{eqn:leader_decision_strategy}. To account for errors of the models \eqref{eqn:follower_decision_strategy} and \eqref{eqn:leader_decision_strategy} from real-world drivers as well as other disturbances, we add a Gaussian noise in \eqref{eqn:compact_dynamics}, leading~to:
\begin{equation}\label{eqn:cut_in_veh_motion_uncertainty}
    \xx(t+1) = \ff\left(\xx(t), u^{0}(t), u^{1, \ast}\left(\xx(t)\right)\right) + w, \,\, w \sim \mathcal{N}(0, \mathcal{W}).
\end{equation}
The cut-in vehicle's action is given~by 
\begin{equation}\label{eqn:deterministic_optimal_action}
u^{1, \ast}\left(\xx(t)\right) = u^{\ast}_{\sigma}\left(\xx(t)\right), \quad \sigma \in \Upsilon = \{\rm l, f\},
\end{equation}
with $u^{\ast}_{\rm l}\left(\xx(t)\right)$ (resp., $u^{\ast}_{\rm f}\left(\xx(t)\right)$) being the first action of the optimal action sequence of the leader, $\uu_{\rm l}^{\ast}(t)$ (resp., of the follower, $\uu_{\rm f}^{\ast}(t)$), determined by \eqref{eqn:leader_decision_strategy} (resp., \eqref{eqn:follower_decision_strategy}). The $\sigma \in \Upsilon = \{\rm l, f\}$ is a latent variable representing the actual role of the cut-in vehicle in the game, which is unknown to the ego vehicle.


We assume that the ego vehicle has a prior belief on $\sigma$, $\bbp(\sigma = \epsilon | \xi(t-1))$. Define the observation history $\xi(t)$ as 
\begin{equation}
    \xi(t) = \{\xx(0), \ldots, \xx(t-1), \xx(t), u^{0}(0), \ldots, u^{0}(t-1)\},
\end{equation}
where $\xx(\cdot)$ are observed traffic states and $u^{0}(\cdot)$ are the actions taken by the ego vehicle at previous times. Then, under the assumption that the cut-in vehicle's role $\sigma$ does not change over time, the ego vehicle can compute a posterior belief on~$\sigma$ according to Bayesian filtering \cite{hwang2006state} as follows:
\begin{equation}\label{eqn:role_prop_update}
    \bbp(\sigma = \epsilon | \xi(t)) \propto \bbp(\xx(t)|\sigma = \epsilon, u^{0}(t-1)) \bbp(\sigma = \epsilon | \xi(t-1)),
\end{equation}
where $\propto$ indicates ``proportional to,'' and the ``likelihood'' function $\bbp(\xx(t)|\sigma = \epsilon, u^{0}(t-1))$ is given as
\begin{equation}
    \bbp(\xx(t)|\sigma = \epsilon, u^{0}(t-1)) = \mathcal{N}(r(t|\epsilon), 0, \mathcal{W}),
\end{equation}
in which $\mathcal{N}(\cdot, 0, \mathcal{W})$ denotes the probability density function of the multivariate normal distribution with zero mean and covariance $\mathcal{W}$ evaluated at $(\cdot)$, and $r(t|\epsilon)$ is the residual between observed state and predicted state assuming the cut-in vehicle's role is $\sigma = \epsilon$, given as
\begin{equation}\label{eqn:prediction_residue}
    r(t|\epsilon) = \xx(t) - \ff\left(\xx(t-1), u^{0}(t-1), u_{\epsilon}^{1, \ast}\left(\xx(t-1)\right)\right).
\end{equation}


At each time $t$, the eco-driving controller uses the computed posterior belief $\bbp(\sigma = \epsilon | \xi(t))$ as the probability distribution of $\sigma$, $\bbp(\sigma)$, in the optimization problem \eqref{eqn:mpc_discrete}. In particular, for each role $\sigma \in \Upsilon = \{\rm l, f\}$, we use either \eqref{eqn:follower_decision_strategy} or \eqref{eqn:leader_decision_strategy} to determine an action sequence of the cut-in vehicle (c), $\uu_{\sigma}^{\rm c}(t) = \{u_{\sigma}^{\rm c}(k|t)\}_{k = 0}^{N-1}$, and then can use the dynamics model~\eqref{eqn:vehicle_kinematics} and $\uu_{\sigma}^{\rm c}(t)$ to obtain a deterministic prediction of the cut-in vehicle's states $\hat{\xx}_{\sigma}^{\rm c}(k |t)$ over the horizon $k = 1,...,N$. This way, the distribution of predicted states $\hat{\xx}_{\sigma}^{\rm c}(k |t)$ is entirely determined by the distribution of $\sigma$, i.e.,
\begin{equation}\label{eqn:cut_in_prediction_prob}
    \hat{\xx}_{\sigma}^{\rm c}(k |t) \sim \bbp(\sigma = \epsilon | \xi(t)), 
    \quad k = 1,\ldots, N.
\end{equation}

\subsection{Prediction on Preceding Vehicles that Accounts for Cut-in Vehicles}\label{sec:prediction_accouts_for_cut_in}
The eco-driving controller \eqref{eqn:mpc_discrete} uses the prediction of the motion of preceding vehicles. 
Besides the prediction on the preceding vehicle, with the game-theoretic model for the cut-in vehicle, we also get the prediction of its motion. We need to fuse the predictions properly before we can solve \eqref{eqn:mpc_discrete} for the eco-driving control. 

For the current non-cut-in (nc) preceding vehicle, we assume it maintains its current speed and lane position in the future~\cite{chaozhe2020fuel}:
\begin{equation}\label{eqn:prediction_no_cut_in}
\begin{aligned}
\hat{\xx}^{\rm nc}(k+1|t) & = f(\hat{\xx}^{\rm nc}(k|t),\, [0, 0]^{\top}),\,  k=0,\ldots, N-1, \\
\Rightarrow \, & \hat{\xx}^{\rm nc} (k|t), \quad k=1,\ldots, N.
\end{aligned}
\end{equation}
This prediction is deterministic.

For the cut-in vehicle (c), based on the estimation of its intended role, we can get the prediction with probabilities~\eqref{eqn:cut_in_prediction_prob}. 
However, the cut-in intention can be to cut in front of or behind the ego vehicle. 
If the cut-in intention is to go behind the ego vehicle, the ego vehicle does not need to react to this cut-in vehicle, especially from an energy efficiency perspective.
To differentiate these two cases, the proposed eco-driving controller first solves \eqref{eqn:mpc_discrete} assuming no cut-in (nc) with the prediction of the current preceding vehicle \eqref{eqn:prediction_no_cut_in}, and get the ego's planned position $s^{\rm 0, nc}(k|t)$ and $l^{\rm 0, nc}(k|t)$.
For each prediction of the cut-in vehicle associated with $\sigma\in \Upsilon$, calculate the crossing step $k^{\text{cut-in}}_{\sigma}$ as the first step when the cut-in vehicle crosses the lane boundary, that is,
\begin{equation}\label{eqn:crossing_step}
\begin{aligned}
    &|l^{\rm c}_{\sigma}(k|t) - l^{0}(k|t)| > \frac{W_{\rm lane}}{2}, \quad \forall \, k < k^{\text{cut-in}}_{\sigma},\\
    &|l^{\rm c}_{\sigma}(k|t) - l^{0}(k|t)| \leq \frac{W_{\rm lane}}{2}, \quad k = k^{\text{cut-in}}_{\sigma}.
\end{aligned}
\end{equation}

\begin{algorithm}
\For{$t=0$ \emph{\KwTo} $t_{\rm f}$}
{Acquire prediction of the current non-cut-in preceding vehicle $\hat{s}^{\rm nc}\big(k|t\big), \hat{v}_{s}^{\rm nc}\big(k|t\big)$ using \eqref{eqn:prediction_no_cut_in}
Solve \eqref{eqn:mpc_discrete} with prediction of non-cut-in preceding vehicle
and get ego's planned position $s^{\rm 0, nc}(k|t)$, $l^{\rm 0, nc}(k|t)$, and desired acceleration $a^{\rm 0, nc}_{\rm d}(q|t)$\;
Observe the traffic states and estimate the role of the potential cut-in vehicle in the leader-follower game using \eqref{eqn:role_prop_update}\;
Acquire prediction of the potential cut-in vehicle $\hat{s}^{\rm c}\big(k|t\big), \hat{v}_{s}^{\rm c}\big(k|t\big)$ using \eqref{eqn:cut_in_prediction_prob}\;
Calculate $k^{\text{cut-in}}_{\sigma}$ according to \eqref{eqn:crossing_step} for each $\sigma \in \Upsilon$, then compute $\Sigma$ using \eqref{eqn:subset_roles}\;
\eIf{$\Sigma = \O$}
{Skip the cut-in vehicle and set $a_{\rm d}(t) = a^{\rm 0, nc}_{\rm d}(q|t)$\;}
{With $\sigma \in \Sigma$ and fused prediction \eqref{eqn:fused_prediction}, solve \eqref{eqn:mpc_discrete} for planned desired acceleration $a^{\rm 0, c}_{\rm d}(q|t)$, and set
$a_{\rm d}(t) = a^{\rm 0, c}_{\rm d}(q|t)$\;}
Apply desired acceleration $a_{\rm d}(t)$ with compensation \eqref{eqn:ucomp_ftil} and 
update ego's states with \eqref{eqn:vehicle_kinematics}\;
}
\caption{Eco-driving controller that accounts for cut-in vehicles}
\label{alg:eco_driving}
\end{algorithm}

With the crossing step identified for each future, a subset of $\sigma$, denoted as $\Sigma \subset \Upsilon$ is computed, which is defined as 
\begin{equation}\label{eqn:subset_roles}
\Sigma = \left\{\sigma | s_{\sigma}^{\rm c}(k|t)-s^{\rm 0, nc}(k|t)\geq \delta s, \exists \,k\geq k^{\text{cut-in}}_{\sigma}\right\}.
\end{equation}
Here $\Sigma$ corresponds to the subset of cut-in vehicle's roles with futures where the cut-in vehicle will cut in front of the ego vehicle if the ego maintains the eco-driving action in reaction to the current preceding vehicle.
Particularly, $\delta s$ is a tuning parameter on how conservative one wants to consider a potential cut-in vehicle.
If $\Sigma =\O$, then in all predicted futures the cut-in vehicle will cut behind the ego vehicle. In this case, the ego vehicle will skip this cut-in vehicle and apply the control action calculated for the current preceding vehicle. 
If $\Sigma \neq\O$, then the cut-in vehicle may cut in front of the ego vehicle. 
Then for each $\sigma \in \Sigma$, we acquire the following prediction on the preceding vehicle
\begin{equation}\label{eqn:fused_prediction}
\hat{\xx}_{\sigma}^{\rm P}(k|t) = 
\begin{cases}
\hat{\xx}^{\rm nc}(k|t)    & {\rm if }\, k < k^{\text{cut-in}}_{\sigma}, \\
\hat{\xx}_{\sigma}^{\rm c}(k |t) \sim \bbp(\sigma = \epsilon | \xi(t))& {\rm if }\, k \geq k^{\text{cut-in}}_{\sigma},
\end{cases}  
\end{equation}
where $k^{\text{cut-in}}_{\sigma}$ is the crossing step defined by \eqref{eqn:crossing_step}.
Note that at $k=k^{\text{cut-in}}_{\sigma}$, there will be a sudden change on $\hat{\xx}^{\rm P}$ from the one that corresponds to the original preceding vehicle to one that corresponds to the cut-in vehicle.

Combining all the steps in this section, the eco-driving controller proposed in this paper is summarized in Algorithm\,\ref{alg:eco_driving}.

\section{Simulation Case Study}\label{sec:results}
In this section, we present a simulation case study for the scenario illustrated in Fig.\,\ref{fig:setup} to demonstrate the effectiveness of the proposed eco-driving controller when handling cut-in vehicles.
We introduce the setup with the parameters used, the energy consumption metric, and the baselines used before we present the simulation results.

\subsection{Simulation Setup}
For the traffic condition visualized in Fig.\,\ref{fig:setup}, we consider parameter values and initial values summarized in Table\,\ref{tab:scenario}, and vary the initial position of the cut-in vehicle $s^{1}(0)$. 

For the leader-follower model on the cut-in vehicle, the parameter values are summarized in Table\,\ref{tab:game_param} and we apply a few simplification steps.
Firstly, consider subsets of the action sequence based on the interactive nature of the cut-in game. 
For traveling straight actions, consider mild actions only: $A_{\rm straight} =$\{\textit{Maintain, Mildly accelerate, Mildly decelerate}\} to mimic a natural mild approaching to slow the preceding vehicle. 
For lane change, we consider hard actions only: $u \in A_{\rm lc} =$\{\textit{Maintain, Hard accelerate, Hard decelerate, Steer to right}\}, with consecutive lateral actions that finish the cut-in motions. 
For lane change abort sequence, we consider only hard decelerate actions $A_{\rm lc} = $\{\textit{Maintain, Hard decelerate, Steer to left}\} and consecutive lateral actions that move the vehicle back to its original lane.
Secondly, while the time horizon $T$ is the same for both the leader-follower game-theoretic model of the cut-in vehicle and the eco-driving controller for the ego, we use a time step $\Delta t = 1$[s], larger than the simulation step, for the leader-follower game-theoretic model to reduce the decision space size. 
Thirdly, once the cut-in vehicle reaches the target lane with $|l-l_{\rm target}| < \delta l$, it is considered that the cut-in is finished, and the cut-in vehicle will start using a car-following model \eqref{eqn:OVM} (introduced later) to determine its acceleration when following its new preceding vehicle.
With these simplifications, we reduced the computation complexity for \eqref{eqn:follower_decision_strategy}
and \eqref{eqn:leader_decision_strategy}. 
The game-theoretic model by the cut-in vehicle is run at a frequency of 2Hz, and the action between runs is determined by a zero-order hold.

For the proposed eco-driving controller in Algorithm~\ref{alg:eco_driving}, the parameter values used are summarized in Table~\ref{tab:mpc_param}. The parameter values are selected based on typical values used in the literature, e.g., \cite{Shen2023ReactiveVsPredictive}, \cite{Li2020GameOriginal}.
The optimization problem~\eqref{eqn:mpc_discrete} is implemented in MATLAB 2023a using YALMIP~\cite{yalmip} and solved with Gurobi~\cite{gurobi}.


\subsection{Energy Consumption Metric}
To account for different powertrain configurations (e.g., internal combustion engine or electric motor), we can use the following metrics, energy consumption per unit mass,
\begin{equation}\label{eqn:w_def}
w = \int_{t_0}^{t_{\rm f}} v_{s}(t)g\Big(a_{s}(t) + \varrho(v_{s}(t))\Big)\mathrm{d}t,
\end{equation}
where $g(\cdot)=\max\{\cdot, 0\}$ implies that braking does not consume or recover energy. We remark that the effects of energy-recovering systems can be included by choosing different $g$ functions.
In simulations, the nonlinear physical term $\varrho$ in\,\eqref{eqn:simple_nonlin_dyn} and \eqref{eqn:w_def} is set to
\begin{equation}
\varrho(v_{s}) = 0.0147 + 2.75\times 10^{-4} v_{s}^2,
\end{equation}
which is acquired for a standard passenger vehicle \cite{Shen2023ReactiveVsPredictive}.
Note in \eqref{eqn:mpc_discrete} to generate a convex quadratic objective function, we choose to optimize the desired acceleration $a$ rather than the energy metrics \eqref{eqn:w_def} by dropping the nonlinear term $\varrho$ in \eqref{eqn:w_def}. 
As shown in the simulation results in this paper, such an approximation is suitable for eco-driving controllers to balance optimality and computation efficiency.

\subsection{Baselines}
\begin{figure}
    \centering
    \includegraphics[width=0.48\textwidth]{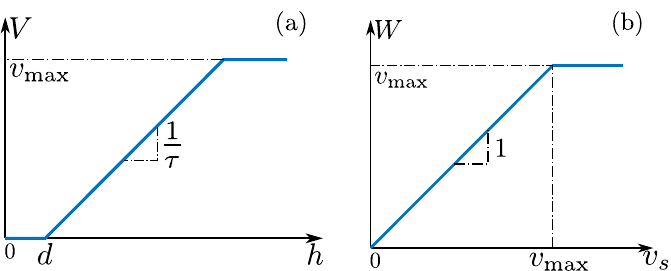}
    \caption{Nonlinear functions in the optimal velocity model\,(OVM). (a) Range policy\,\eqref{eqn:range_policy}. (b) Speed policy\,\eqref{eqn:speed_policy}.}
    \label{fig:ovm_fun}
    \vspace{-4mm}
\end{figure}

We use two baselines for the proposed eco-driving controller: one car-following model and one eco-driving controller that does not account for cut-in vehicles.

The first baseline uses the optimal velocity model\,(OVM) which yields the following desired acceleration \cite{orosz2020survey}: 
\begin{equation}\label{eqn:OVM}
a_{\rm d}^{\rm OVM} = \alpha \big(V(h) - v_{s}\big) + \beta \big(W(v^{\rm P}_{s}) - v_{s}\big).
\end{equation}
Here $v_{s}$ and $v^{\rm P}_{s}$ is the longitudinal speed of the ego vehicle and its preceding vehicle in the same lane. $V(h)$ is the range policy that determines the desired velocity as a function of the distance headway\,$h = s^{\rm P} - s - L_{\rm veh}$ with $s$ and $s^{\rm P}$ being the longitudinal positions. Consistent with eco-driving in \eqref{eqn:constant_time_headway}, we use the following constant time headway range policy:
\begin{equation}\label{eqn:range_policy}
V(h) = \min\Big\{v_{\max}, \max\big\{0, (h-d) / \tau\big\}\Big\}.
\end{equation}
As is shown in Fig.\,\ref{fig:ovm_fun}(a), when the distance headway is less than the stopping distance $d$, the ego vehicle tends to stay still, while when the distance headway is larger than $d+ \tau v_{\max}$, the ego vehicle intends to travel with maximum speed $v_{\max}$ without being influenced by the preceding vehicle. 
Moreover, the speed policy 
\begin{equation}\label{eqn:speed_policy}
W(v^{\rm P}_{s}) = \min\big\{v_{\max},v^{\rm P}_{s}\big\},
\end{equation}
is used to prevent the ego vehicle from speeding once the preceding vehicle goes faster than $v_{\max}$; see Fig.\,\ref{fig:ovm_fun}(b).
This baseline is referred to as ``OVM" in this section.

The second baseline is the eco-driving controller which does not account for cut-in vehicles. This is achieved by only run steps 2, 7, and 11 in Algorithm \ref{alg:eco_driving}. 
This baseline is referred to as ``Eco-driving" in this section.

For both baseline controllers, there is no special consideration with regard to cut-ins except for that they keep monitoring the preceding vehicle using \eqref{eqn:determine_preceding_vehicle}. When a cut-in vehicle enters ego's lane, there will be a sudden change in the lead vehicle's states and both controllers react to this cut-in vehicle as a new preceding vehicle.

\subsection{Eco-driving Controller Validation: No Cut-in Vehicle}
We first consider a traditional no cut-in vehicle scenario to confirm the benefit of the eco-driving controller in traffic with the baseline controllers. 
We limit the simulation to the single lane where the ego is approaching the slow traffic flow ahead. We simulated three cases to also highlight the impact of delay: 1) there is no delay in the powertrain $\iota = 0.0$[s]; 2) there is delay $\iota = 0.6$ [s] and the baseline controllers consider the delay; 3) there is delay $\iota = 0.6$ [s] but the baseline controllers do not consider the delay.

The energy consumption for all the cases is summarized in Table\,\ref{tab:energy_results_no_cut_in}. For all cases, the baseline eco-driving controller indeed achieved better energy consumption compared to OVM baseline. 
Furthermore, the presence of time delay increases energy consumption for both controllers. While the OVM design does not alter its behavior as it does not explicitly consider the time delay, the MPC-based eco-driving controller, without considering the delay leads to a noticeable increase in energy consumption.
\begin{table}[h]
    \centering
    \caption{Energy consumption when no cut-in vehicle. }
    \begin{tabular}{|c|c|c|c|}
       \hline 
       \multirow{2}{*}{\diagbox[width=10em]{Type}{Energy  $\mathrm{[J/kg]}$}}  & No delay & \textbf{With delay} & With delay \\ 
       & $\iota = 0.0$[s]  & \textbf{$\boldsymbol{\iota = 0.6}$[s]} & but not consider \\ \hline
       OVM  & 86.98 & \textbf{93.72} & 93.72 \\ \hline
       Eco-driving  & 80.79 & \textbf{82.17} & 82.46 \\ \hline
    \end{tabular}
    \label{tab:energy_results_no_cut_in}
\end{table}

\begin{figure}[h]
    \centering
    \includegraphics[width=0.48\textwidth]{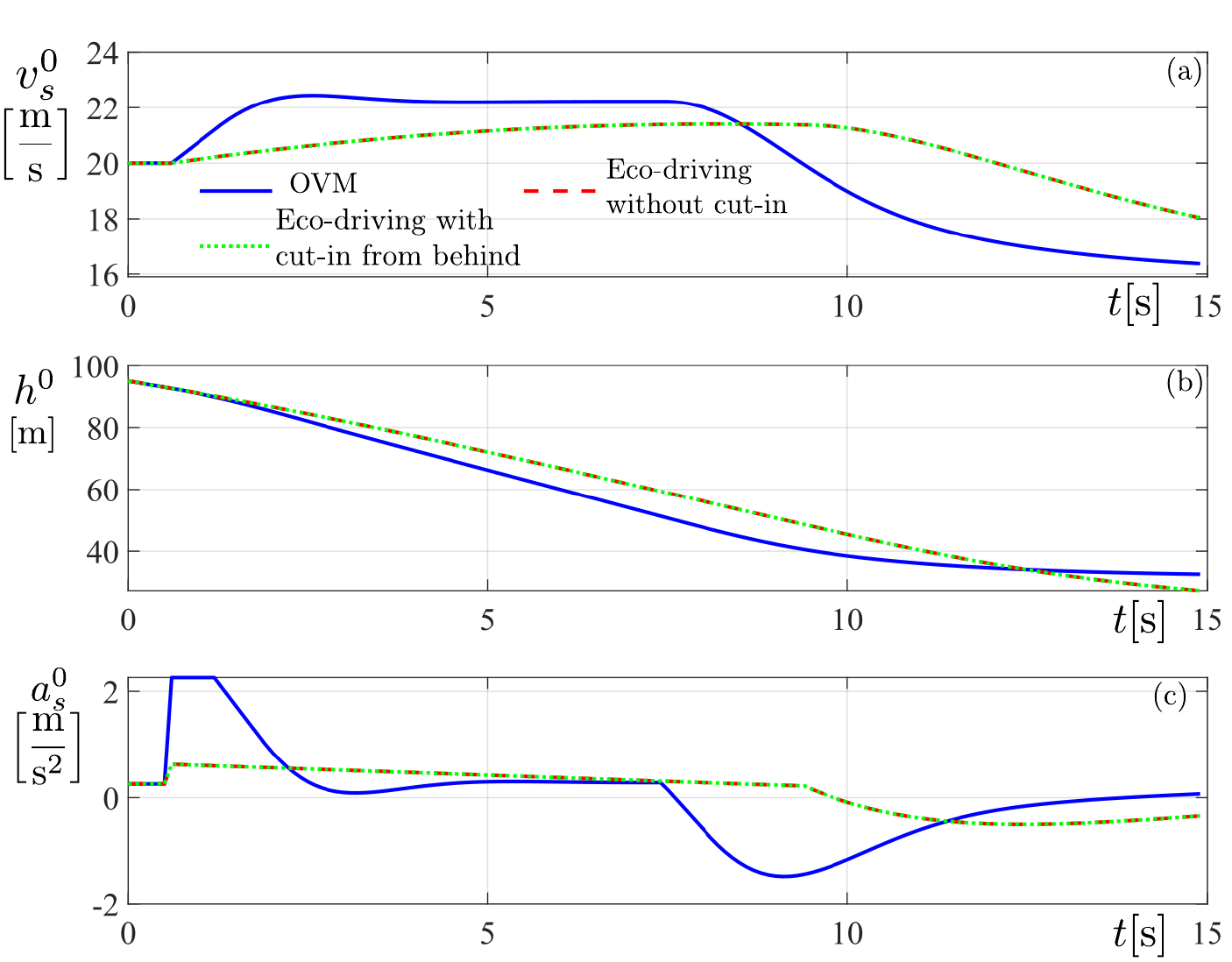}
    \caption{The time profile of the ego vehicle with different controllers and scenarios (a) longitudinal speed (b) distance headway (c) longitudinal acceleration. In all panels, the blue solid curves correspond to the case when the ego is using OVM controller \eqref{eqn:OVM} and no vehicle cut-in from the adjacent lane; the red dashed curves correspond to the case when the ego is using baseline eco-driving controller with no vehicle cut-in from the adjacent lane; the green dashed-dotted curves correspond to the case when the ego is using the eco-driving controller in Algorithm\,\eqref{alg:eco_driving} with a vehicle cut-in from behind from the adjacent lane, and the first 4 seconds top view of this scenario is shown in Fig.\,\ref{fig:cut_in_from_behind_top_view}.}
    \label{fig:no_cutin_vs_cut_in_from_behind}
\end{figure}

In Fig.\,\ref{fig:no_cutin_vs_cut_in_from_behind}, the time profile of longitudinal speed $v_{s}$, distance headway $h^{0}$ and the longitudinal acceleration $a_{s}$ for the ego vehicle are plotted, corresponding to the case where powertrain delay $\iota = 0.6$\,[s] and it is considered by the eco-driving controller. 
In all three panels, the blue solid and red dashed curves correspond to those of the OVM controller and eco-driving controller respectively. As intended, the eco-driving controller commands less aggressive acceleration at the first part and avoids hard brakes while staying safe and approaching the slow traffic ahead. 
\begin{figure}[h]
    \centering
    \includegraphics[width=0.48\textwidth]{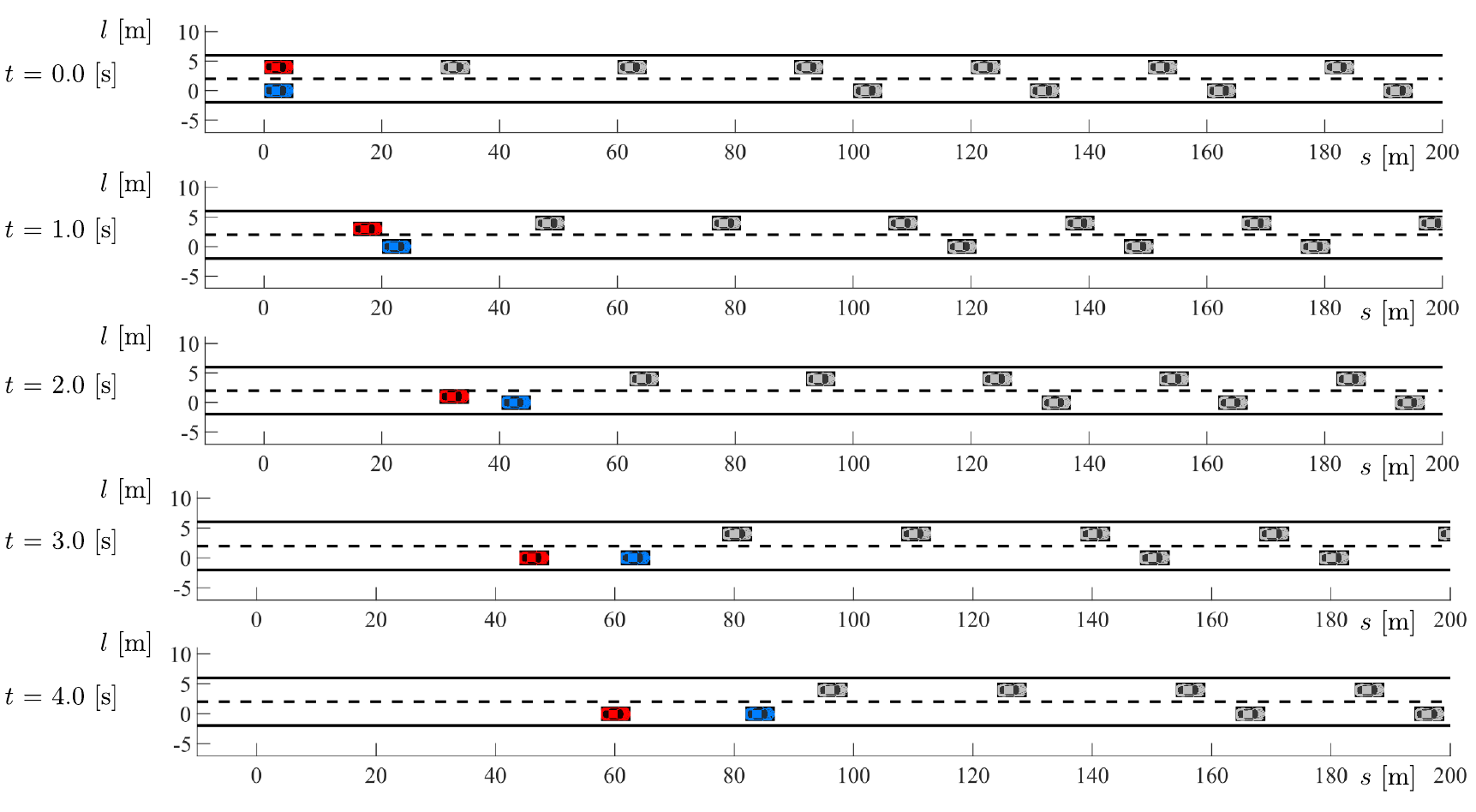}
    \caption{Top view of the scenario when the cut-in vehicle cut from the behind of the ego vehicle, with the screenshot corresponding to every second up to the 4 seconds, which covers the whole cut-in procedure. 
    In all screenshots, the blue vehicle is the ego, the red vehicle is the cut-in vehicle and the grey vehicles are other vehicles traveling in the slow traffic ahead. The ego (blue vehicle) motion corresponds to the first 4 seconds of the green dashed-dotted profiles in Fig.\,\ref{fig:no_cutin_vs_cut_in_from_behind}.}
    \label{fig:cut_in_from_behind_top_view}
\end{figure}

For the remainder of this section, we fix the powertrain delay to $\iota = 0.6 $ [s] in all simulations and consider it in the eco-driving controller.

\begin{figure*}
    \centering
    \includegraphics[width=0.99\textwidth]{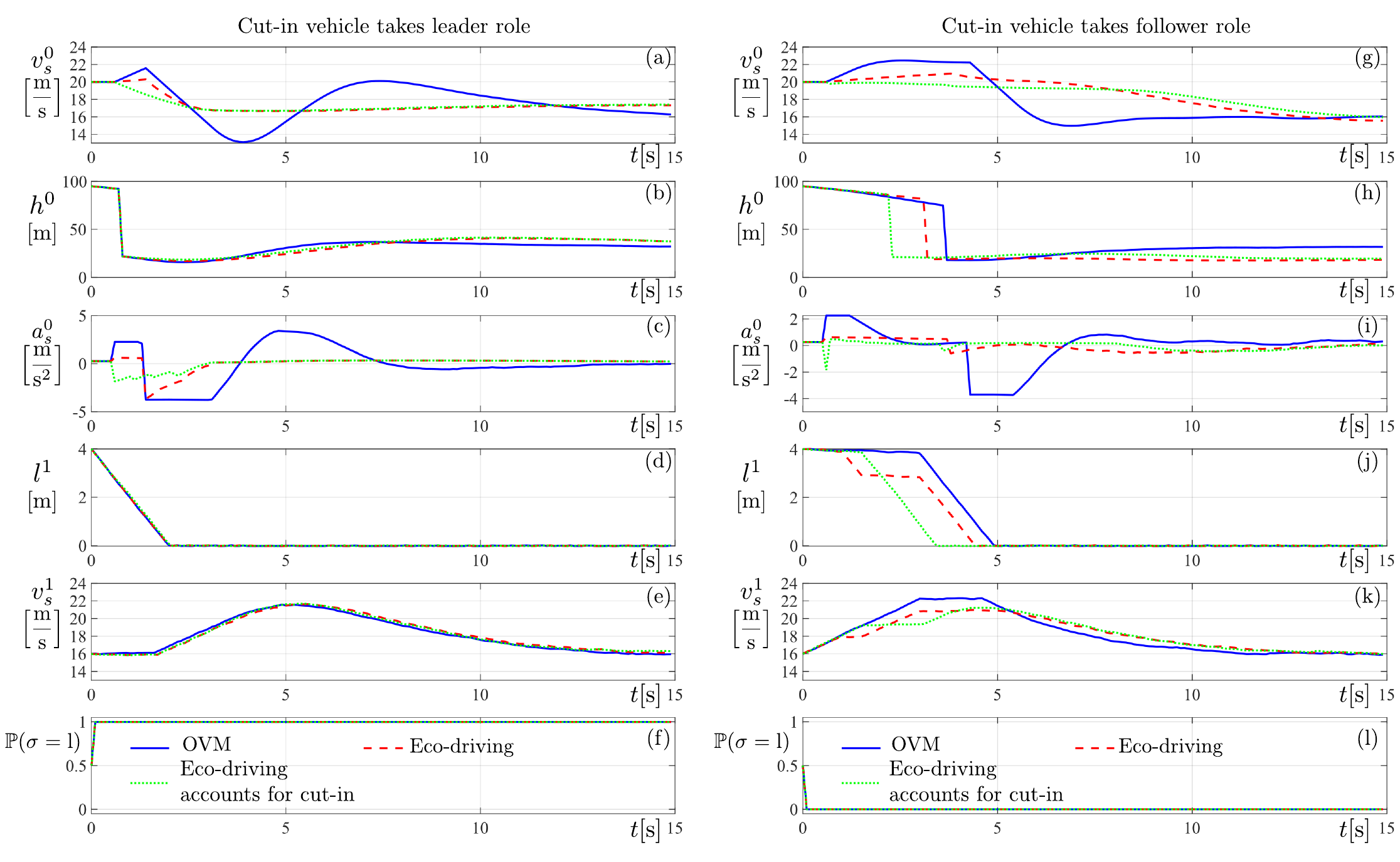}
    \caption{The time profiles of the ego: (a) (g) longitudinal speed $v^{0}_{s}$, (b) (h) distance headway $h^{0}$, (c)(i) longitudinal acceleration $a_{s}^{0}$), the cut-in vehicles ((d)(j) lateral position $l^{1}$, (c)(k) longitudinal speed $v_{s}^{1}$, and (f)(l) the estimated probability of the cut-in vehicle's role being the leader. 
    The left and right columns are for the case when the cut-in vehicle takes the leader role and the follower role, respectively.
    In all panels, the blue solid curves correspond to the run when the ego is using the baseline OVM controller \eqref{eqn:OVM}, the red dashed curves correspond to the run when the ego is using the baseline eco-driving controller, while the green dashed-dotted curves correspond to the run where the ego is using the proposed eco-driving controller Algorithm\,\ref{alg:eco_driving}.}
    \label{fig:normal_cut_in}
\end{figure*}

\begin{figure*}[t]
    \centering
    \includegraphics[width=0.99\textwidth]{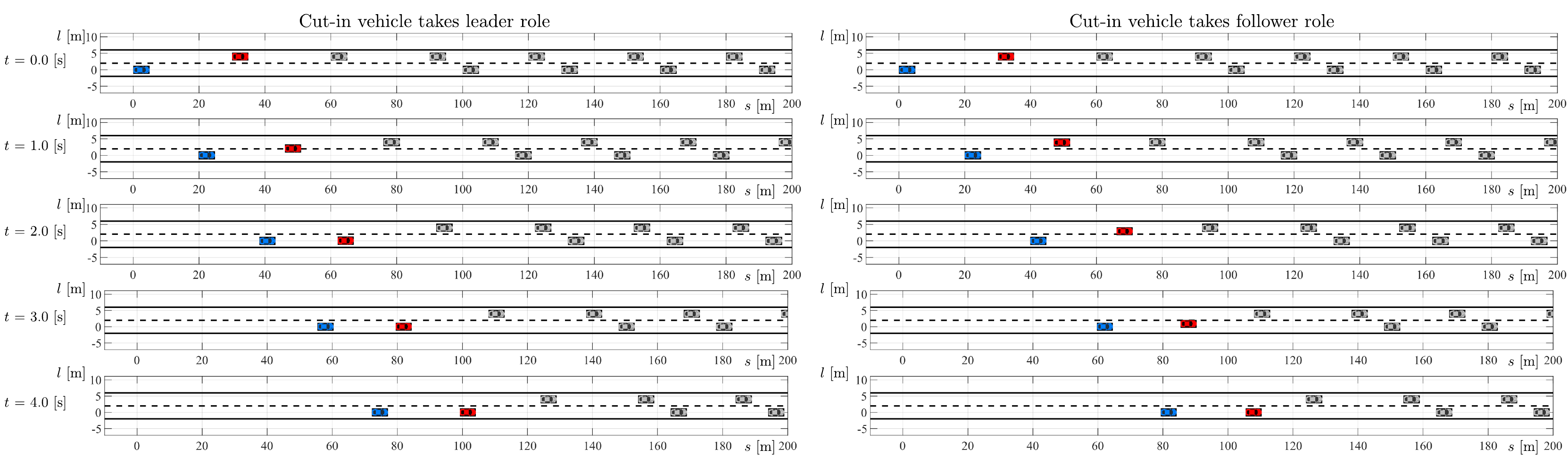}
    \caption{Top view of the scenario when the cut-in vehicle cut in front of the ego, with the screenshot corresponding to every second up to the 4 seconds that cover the whole cut-in procedure. The left and right columns correspond to the cases when the cut-in vehicle takes the leader role and the follower role, respectively.
    In all screenshots, the blue vehicle is the ego, the red vehicle is the cut-in vehicle and the grey vehicles are other vehicles traveling in the slow traffic ahead. The ego (blue vehicle) motion corresponds to the first 4 seconds of the green dashed-dotted profiles in Fig.\,\ref{fig:normal_cut_in}.}
    \label{fig:normal_cut_in_top_view}
\end{figure*}
\subsection{Vehicle Cut-in from Behind}
Next, we consider a scenario when the cut-in vehicle will cut from behind the ego vehicles, by setting $s^{1}(0) = 0$ [m], that is, the potential cut-in vehicle is parallel to the ego vehicle initially. 
The potential cut-in vehicle decides to cut in from behind for both leader and follower roles. 
The first 4 seconds top view of the scenario is shown in Fig.\,\ref{fig:cut_in_from_behind_top_view}, where the blue vehicle is the ego, the red vehicle is the cut-in vehicle and the grey vehicles are other vehicles traveling in the slow traffic ahead. The time profiles of the ego vehicle with the eco-driving controller that considers this cut-in vehicle are plotted in Fig.\,\ref{fig:no_cutin_vs_cut_in_from_behind} panels as green dashed-dotted curves. 
The proposed controller determines that this cut-in vehicle will cut in from behind and ignores it, as indicated by the fact that the red dashed curves and the green dashed-dotted curves overlap with each other in all panels in Fig.\,\ref{fig:no_cutin_vs_cut_in_from_behind}.
On the other hand, the interactive nature of the leader-follower model is demonstrated: to achieve the cut-in, it makes a tactical decision to slow down and then cut to the ego lane.

\subsection{Vehicle Cut-in in the Front}
Finally, we present a scenario when the cut-in vehicle cuts in front of the ego vehicles, by setting $s^{1}(0) = 30$\,[m]. The cut-in vehicle decides to cut in front of the ego vehicle but depending on the roles, and the ego vehicle's behavior, the cut-in vehicle could behave differently. 

We ran simulations with the cut-in vehicle taking different roles. Given the uncertainty on the cut-in vehicle's action \eqref{eqn:cut_in_veh_motion_uncertainty}, we run the same scenario 10 times for each controller and each role and summarized the energy consumption in Table.\,\ref{tab:energy_results_with_cut_in}.
This proposed controller achieved consistent major reductions in energy consumption compared to baselines: 10.5\% over baseline eco-driving controller and 68.4 \% over baseline OVM controller for a cut-in vehicle that takes the leader role while 32.2\% over baseline eco-driving controller and 76.1 \% over baseline OVM controller for a cut-in vehicle that takes the follower role. 

To compare the motions by different controllers, the time profiles of the ego (longitudinal speed $v^{0}_{s}$, distance headway $h^{0}$ and longitudinal acceleration $a_{s}^{0}$) and the cut-in vehicles (lateral position $l^{1}$ and longitudinal speed $v_{s}^{1}$) that corresponds to one of the simulation tuples are plotted in Fig.\,\ref{fig:normal_cut_in} (a)-(e) for the case when the cut-in vehicle take the leader role, and in Fig.\,\ref{fig:normal_cut_in} (g)-(k) for the follower role. 
The estimation of the cut-in vehicle's role is plotted in Fig.\,\ref{fig:normal_cut_in} (f) and Fig.\,\ref{fig:normal_cut_in} (l) accordingly. 
In all panels, the blue solid curves correspond to the run when the ego is using the baseline OVM controller \eqref{eqn:OVM}, the red dashed curves correspond to the run when the ego is using the baseline eco-driving controller, while the green dashed-dotted curves correspond to the run where the ego is using the proposed eco-driving controller in Algorithm\,\ref{alg:eco_driving}. 
As shown in Fig.\,\ref{fig:normal_cut_in} (f) and (l), the estimated role of the cut-in vehicle quickly converges to the actual one, demonstrating the effectiveness of the estimation method \eqref{eqn:role_prop_update}.
On the other hand, while all controllers can drive the ego approaching slow traffic and handle the cut-in vehicle safely, the proposed design uses the least brakes and accelerations with the proper prediction of the cut-in vehicle's motion, which explains the improvements in Table \ref{tab:energy_results_with_cut_in}.

To further compare the different interactive behaviors established by the cut-in vehicle when taking different roles as well as when the ego establishes different behaviors, we plot the top view screenshots of the first 4 seconds corresponding to the case when the ego is running the proposed eco-driving controller in Fig.\,\ref{fig:normal_cut_in_top_view} for the cut-in vehicle takes leader role (left column) and follower role (right column) respectively.
When taking the leader role, the cut-in vehicle cuts into the ego's lane sooner than when it holds follower roles. 
Indeed, the leader-follower game-theoretic model can reflect different aggressiveness by the drivers by taking different roles. Despite the different behavior of the ego vehicle in reaction to the cut-in vehicles, the cut-in vehicle that takes the leader role does not change its behavior. This is reasonable considering its decision process is \eqref{eqn:leader_decision_strategy} is assuming that the follower will take the ``max-min" solution in reaction to its motion, and the most rewarding action for itself is to cut in front of ego as soon as possible and acceleration to catch the traffic ahead.  
By contrast, when the cut-in vehicle takes the follower role, its cut-in behavior changes in reaction to the ego's behavior.
As shown in Fig.\,\ref{fig:normal_cut_in}\,(j) and (k), when the ego is running with the proposed eco-driving controller or baseline eco-driving controllers, the cut-in vehicle cut-in earlier and did not accelerate to high speed, as opposed to its behavior in reaction to ego's motion when running with baseline OVM. 
Such differences in behavior demonstrate the capability of the leader-follower game in modeling interaction between vehicles. 
More importantly, despite behavior changes by the cut-in vehicle in reaction to our eco-driving action, good energy efficiency is still obtained, which demonstrates the effectiveness of the proposed eco-driving controller in interactive scenarios.

\begin{table}
    \centering
    \caption{Energy consumption when a cut-in vehicle takes different roles.}
    \begin{tabular}{|c|c|c|}
       \hline 
       \multirow{2}{*}{\diagbox[width=14em]{Type}{Energy $\mathrm{[J/kg]}$}}  & Leader role & Follower role 
       \\ 
       & (mean / std) &(mean / std) 
       \\ \hline
       OVM  & 189.06 / 21.68 & 118.34 / 2.79\\ \hline
       Eco-driving  & 66.73 / 1.32 & 41.79 / 0.53\\ \hline
       Eco-driving accounts for cut-in & \textbf{59.70 / 0.78} & \textbf{28.32 / 0.65} \\ \hline
    \end{tabular}
    \label{tab:energy_results_with_cut_in}
\end{table}

\section{Conclusions}\label{sec:concl}

In this paper, an eco-driving controller that accounts for interactive cut-in vehicles has been proposed. A leader-follower game-theoretic model has been utilized to represent the cut-in vehicle's decision process. With such a model, an MPC-based eco-driving controller has been designed which can predict the cut-in vehicle's behavior and generate corresponding energy-efficient motion plans for the ego vehicle. Simulation case studies have been presented to demonstrate the effectiveness of the proposed controller over baselines that do not account for cut-in vehicles over prediction. Future research includes validating the proposed controller with higher-fidelity simulations, using real-world traffic datasets, and/or on a real vehicle to evaluate its benefits in more realistic environments.

\bibliographystyle{IEEEtran}
\bibliography{ref.bib}
\appendices
\newpage
\section{Summary on Parameter Values}\label{sec:appdx_param}

\begin{table}[h]
    \centering
    \caption{Traffic Scenario Parameters}
    \begin{tabular}{cccc}
     $L_{\rm veh}$ [m]& $W_{\rm veh}$ [m]& $W_{\rm lane}$ [m] & $l_{\rm target}$ \\ \hline
     5.0 & 2.5 & 4.0 &  0.0 \\ 
     \rule{0pt}{10pt}$v_{\max}\,{\rm [m/s]}$ & $\Delta t$ [s] &  $ t_{\rm f}$ [s] & $\delta l$ [m] \\ \hline
     30.0 & 0.1 &  15.0 & 1.0 \\
    \rule{0pt}{10pt} $s^{0}(0)$ [m]& $h^{0}(0)$ & $h^{1}(0)$ [m]& $v^{0}(0)\,{\rm [m/s]}$   \\ \hline
    0.0 & 95.0 & 25.0 & 20.0 \rule{0pt}{10pt} \\
    \multicolumn{2}{c}{ $v^{i}(0)\,{\rm [m/s]},\,i\geq 1$} &
    \multicolumn{2}{c}{ $\mathcal{W}$} \\ \hline
    \multicolumn{2}{c}{16.0} &
    \multicolumn{2}{c}{\text{diag}(0.002, 0.001, 0.0002)}\\
    \end{tabular}%
    \label{tab:scenario}
\vspace{-4mm}
\end{table}

\begin{table}[h]
    \centering
    \caption{Leader-follower game model parameters}
    \begin{tabular}{cccc}
     $\tau_{\rm desired}$ [s]& $a_{\text{mild}}\,{\rm [m/s^2]}$ & $a_{\text{hard}}\,{\rm [m/s^2]}$ & $\lambda$ \\ \hline
    1.0 &1.33& 2.0 & 0.9 \rule{0pt}{10pt} \\ 
    $\Delta t$\, [s] & $N$ & \multicolumn{2}{c}{$\boldsymbol{\omega}$ } \\ \hline 
    1 & 5 & \multicolumn{2}{c}{[400, 5, 1, 40, 0, 0.1]}
    \end{tabular}%
    \label{tab:game_param}
\vspace{-4mm}
\end{table}

\begin{table}[h!]
\centering
\caption{Eco-driving controller parameters (including baselines)}
\begin{tabular}{ccccccc}
$\Delta t\,\mathrm{[s]}$ & $q_{\mathrm{g}}$ & $q_{\mathrm{a}}$  & $\tau\,\mathrm{[s]}$  & $d\,\mathrm{[m]}$ & $\tau_{\min}\,\mathrm{[s]}$ & $d_{\min}\,\mathrm{[m]}$   
\\ \hline
0.1     & 1  & 960  & 1.67 & 5  & 0.67  & 3\\
\rule{0pt}{10pt} $T$ [s] & $N$ & $\delta s$ [m] & $\eta$ & $\alpha$ [1/s] & $\beta$ [1/s] &\\ \hline
    5.0 & 50 & 0.0 & 0.03 & 0.4 & 0.5 &
\end{tabular}
\label{tab:mpc_param}
\vspace{-8mm}
\end{table}
\end{document}